# APT-MMF: An advanced persistent threat actor attribution method based on multimodal and multilevel feature fusion

Nan Xiao[a,*], Bo Lang[a,b,*], Ting Wang[a,c], Yikai Chen[a]

[a] State Key Laboratory of Software Development Environment, Beihang University, Beijing, 100191, China
[b] Zhongguancun Laboratory, Beijing, China
[c] National Computer Network Emergency Response Technical Team/Coordination Center of China, Beijing, China
*Corresponding author. E-mail addresses: xiaonan@buaa.edu.cn (N. Xiao), langbo@buaa.edu.cn (B. Lang).

## Abstract

Threat actor attribution is a crucial defense strategy for combating advanced persistent threats (APTs). Cyber threat intelligence (CTI), which involves analyzing multisource heterogeneous data from APTs, plays an important role in APT actor attribution. The current attribution methods extract features from different CTI perspectives and employ machine learning models to classify CTI reports according to their threat actors. However, these methods usually extract only one kind of feature and ignore heterogeneous information, especially the attributes and relations of indicators of compromise (IOCs), which form the core of CTI. To address these problems, we propose an APT actor attribution method based on multimodal and multilevel feature fusion (APT-MMF). First, we leverage a heterogeneous attributed graph to characterize APT reports and their IOC information. Then, we extract and fuse multimodal features, including attribute type features, natural language text features and topological relationship features, to construct comprehensive node representations. Furthermore, we design multilevel heterogeneous graph attention networks to learn the deep hidden features of APT report nodes; these networks integrate IOC type-level, metapath-based neighbor node-level, and metapath semantic-level attention. Utilizing multisource threat intelligence, we construct a heterogeneous attributed graph dataset for verification purposes. The experimental results show that our method not only outperforms the existing methods but also demonstrates its good interpretability for attribution analysis tasks.

## 1. Introduction

Advanced persistent threats (APTs) refer to targeted and continued threats carried out by sophisticated attackers with vast resources[1]. Characterized by high levels of stealthiness, these threats can persist over the long term, resulting in disruptions or other severe consequences. It is difficult to prevent APTs entirely. Skilled attackers have the opportunity to bypass existing defense systems by leveraging technical advantages such as 0-day vulnerabilities[2]. Therefore, it is necessary to research APT attribution to improve the deterrence provided by security defenses[3–6].

APT attribution is a challenging task. Attackers often employ various strategies to obfuscate their tracks and even mislead the attribution process[7]. Defenders must investigate indicators of compromise (IOCs) and assess tactics, techniques, and procedures (TTPs, also known as high-level IOCs) within massive amounts of data; they also need to analyze the traits of attackers and make great efforts to confirm their identities. Early studies proposed a few APT attribution models based on practical experience[3,4,8–12], implementing attribution analysis and inspiring subsequent technical research. According to the analysis goals, APT attribution can be divided into two stages: attack process attribution using attack traces and threat actor attribution using attack features[4,9]. Technical approaches for APT attribution include rule reasoning methods and machine learning methods. Rule reasoning methods involve argument-based reasoning[13–17] and graph rule-based reasoning[18–20]. These methods require numerous rules to be manually designed by experts, and most of the formed rules focus on attack process attribution rather than threat actor attribution. Researchers have recently developed machine learning-based APT attribution methods for attack process attribution[21–26] and threat actor attribution[27–41]. The available methods for threat actor attribution can be divided into malware-based methods and cyber threat intelligence (CTI)-based methods. Malware is an important component of APTs. It usually has various functions, such as command and control and is essential for attribution. However, malware can be reused by different APT groups. Only using malware for threat actor attribution presents significant limitations.

The CTI reports of APTs aggregate various characteristic and analytical information related to APT groups, such as malicious internet protocols (IPs) or domains used for command and control, as well as the tactics and techniques employed in the attack process. These APT reports are usually compiled by cybersecurity experts based on the collection and analysis of multisource data, such as network traffic, malware, and open-source intelligence. Compared to malware, CTI has richer contextual information and more diverse attribution data. Therefore, APT reports are regarded as a crucial data source for attribution analysis[42–45].

APT actor attribution based on CTI has aroused considerable research interest in recent years. The related methods usually extract various features from CTI reports and input them into traditional machine learning models such as support vector machines (SVMs) and multilayer perceptrons (MLPs) to classify CTI reports according to their threat actors. These features can be categorized into three types: **(1) Tactic and technique features**[35–37]. The tactic and technique features are extracted from CTI reports using one-hot encoding[35,36] or term frequency-inverse document frequency (TF-IDF)[37] based on the adversarial tactics, techniques, and common knowledge (ATT&CK) framework. However, the expressivity of one-hot encoding or TF-IDF is limited, and in the available attribution data types exhibit a lack of diversity due to the use of only tactic and technique features. **(2) Text features**[38–40]. The text features of CTI reports are extracted based on self-constructed vocabulary via bag-of-words or Word2Vec models in related methods[38–40]. The expressivity of these text features is constrained by the scale and quality of the given vocabulary or training corpus. In addition, when using only text features, these methods ignore the types and topological information of indicators in threat actor attribution scenarios. **(3) Homogeneous topological features**[41]. Huang et al.[41] used a graph to model relation information in CTI and extracted homogeneous topological features via random walk-based graph embedding. However, using only homogeneous topological features results in the inability to fuse heterogeneous graph information for more comprehensively and accurately performing attribution analysis. In summary, the existing CTI-based methods do not sufficiently extract and fuse various features, particularly ignoring the heterogeneity of CTI information. In addition, few types of IOCs are used, and IOC attributes are often overlooked.

IOCs refer to evidence that the victim's network has been compromised. These indices are the most practical indicators in current APT detection and attribution methods and are the core of CTI reports. IOCs include numerous types, including low-level IOCs such as IPs and domains and high-level IOCs such as tactics and techniques. IOCs often have rich attributes. For example, an IP has an IP address and a geolocation attribute. In addition, various relationships are present among IOCs, such as communication between malware and domains.

Based on the above analysis, we leverage a heterogeneous attributed graph to characterize APT reports and their included IOCs, and we conduct a study on APT actor attribution according to IOCs. We propose an **APT** actor attribution method based on **M**ultimodal and **M**ultilevel feature **F**usion, named APT-MMF. First, we design a heterogeneous attributed graph schema for APT actor attribution based on various CTI models. Our schema centers on an attributeless APT report and is connected to multiple IOC types with various attributes and interrelations. Then, we propose a method for extracting and fusing multimodal node features, including attribute type features, natural language text features, and topological relationship features. Compared to the existing studies, we enhance the attribute type feature extraction process by utilizing diverse node and attribute types in the designed schema, improve the natural language text feature extraction procedure with the semantically powerful and pretrained bidirectional encoders from transformers (BERT) model[46], and employ Node2vec[47] to efficiently extract topological relationship features. Next, to further learn the deep features of APT report nodes, we propose multilevel heterogeneous graph attention networks to fuse heterogeneous graph features with different importance. Specifically, we propose IOC type-level attention, utilizing the IOC neighbors of APT report nodes to complete the sparse features of the attributeless report nodes derived from multimodal extraction and fusion. Based on this level, we integrate metapath-based neighbor node-level and metapath semantic-level attention[48], effectively fusing the information of metapath-based neighbor nodes, which are report-type neighbors, along with semantic metapath information. Finally, we input the learned representations of the APT report nodes into a fully connected layer for classification and obtain threat actor attribution results.

The contributions of our work are summarized as follows:
1. We design a heterogeneous attributed graph schema for APT actor attribution and propose methods for extracting and fusing multimodal node features, including the type features of node attributes, the natural language features of node attribute texts, and the topological relationship features among nodes. These features comprehensively represent the characteristics of IOC nodes and effectively yield improved attribution performance.
2. We propose multilevel heterogeneous graph attention networks to learn the deep hidden features of APT report nodes, which weight and fuse information from IOC neighbor nodes, metapath-based report neighbor nodes, and metapaths in a hierarchical manner. These triple attention mechanisms complement each other, holistically fusing heterogeneous graph information and effectively improving the attribution performance of the model. To the best of our knowledge, this is the first attempt to study APT actor attribution based on heterogeneous graph representation learning.
3. We conduct APT actor attribution experiments on a heterogeneous attributed graph dataset constructed from multisource CTI reports. The experimental results show that our method achieves a $Micro\text{-}F_1$ value of 83.21% and a $Macro\text{-}F_1$ value of 70.51% in the attribution classification task; these values are significantly greater than those of the baseline methods. Moreover, by analyzing the triple attention mechanisms, our proposed method demonstrates its good interpretability for attribution analysis scenarios.

The remainder of the paper is organized as follows. Section 2 summarizes the related work on APT actor attribution. Section 3 presents the overall framework and specific implementation of our method in detail. Section 4 describes the process of constructing the datasets and the experimental settings and reports and analyzes the results of various experiments. Section 5 summarizes our work and draws conclusions.

## 2. Related Work

In this section, we first summarize the current research on APT actor attribution. Then, we introduce the representative CTI modeling and graph representation learning approaches that are relevant to our research.

## 2.1 APT Actor Attribution

APT actor attribution methods can be divided into malware-based methods and CTI-based methods. Malware-based methods extract various features and employ machine learning models to classify malware accord to their APT actors. These malware features include dynamic features and static features. Dynamic features refer to the behavioral characteristics obtained by executing malware in a simulated environment and monitoring its behavior. Qiao et al.[27] extracted six types of API call behavior features and utilized the Jaccard similarity coefficient for classification purposes. Mei et al.[29] extracted various execution action features based on logs and proposed a particle swarm optimization multiclass SVM method to attribute APT actors. The advantages of dynamic analysis include its ability to identify unknown malware and reduce false positives and false negatives. However, its disadvantages are its high operating environment requirements and the associated time costs. Due to not executing the examined malware, static analysis has the advantages of being secure and fast. However, this approach is limited by reverse engineering techniques, which can potentially lead to omissions of critical information. Laurenza et al.[32] extracted static features such as PE headers and obfuscated strings and classified malware based on random forests. Liu et al.[31] extracted and fused features of the control flow graph and functions based on decompilation and used convolutional neural network (CNN) and long short-term memory (LSTM) models to achieve classification. Wang et al.[31] combined dynamic features such as registry operations and static features such as bytecodes to implement attribution classification based on random forests. Although malware is an important artifact for determining APT attribution, it may be developed and used by multiple APT groups. Therefore, using only malware to attribute APT actors has great limitations.

CTI reports of APTs collect various attribution information for APT actor attribution. Compared to malware, CTI has richer contextual information and can make attribution more comprehensive and accurate. The current CTI-based methods for APT actor attribution extract different features, such as tactic and technique features[35–37], text features[38–40] and homogeneous topological features[41], and employ traditional machine learning models, such as SVMs and MLPs, to implement attribution classification. Tactics and techniques are the core characteristics of APTs. Analyzing the tactic and technique features of APT reports with the ATT&CK framework is important for threat actor attribution[35–37]. Shin et al.[36] utilized one-hot vectors of techniques to represent various tactics in ATT&CK and then analyzed the attribution through weighted summation and cosine similarity calculations of these vectors. Noor et al.[35] utilized the latent semantic analysis (LSA) model to determine whether CTI reports contained each of the 188 selected techniques and to form a 188-dimensional one-hot vector; then, traditional machine learning models such as k-nearest neighbors (KNN) were employed for attribution classification. Lee and Choi[37] used TF-IDF to embed the technique sequences of APT reports sorted by the tactic order of ATT&CK and adopted cosine similarity to analyze attack groups. However, when one-hot or TF-IDF vectors are used for feature representation, the deep semantic relationships among tactics and techniques are omitted. In addition, using only tactic and technique features for APT actor attribution may result in the omission of crucial information in other indicators.

Text features can represent the content semantics of CTI. Some studies have introduced word embedding models to learn text representations of threat intelligence for threat actor attribution[38–40]. Perry et al.[39] constructed their own vocabulary and improved the bag-of-words model to vectorize threat intelligence texts. The authors also applied the extreme gradient boosting (XGBoost) model for attribution classification. Irshad et al.[38] and S et al.[40] followed the approach of Perry et al.[39] and used Word2Vec embeddings of words or their similar words to create a feature matrix of a CTI report. Then, they input this matrix into classifiers such as an MLP for threat actor attribution. These methods use self-constructed and unpublished vocabulary lists to extract text features from threat intelligence data. Words in the vocabulary list that are weakly related to threat actor attribution often result in high-dimensional feature redundancy. Moreover, the expressivity of text features is constrained by the size and quality of the utilized vocabulary list or pretrained corpus. In addition, this vectorization approach can represent only text semantics without capturing the type and relation information of attribution artifacts. Topology features can capture relation information in threat intelligence data. Huang et al.[41] utilized a homogeneous graph embedding method based on random walks to extract topological features from threat intelligence data. They also employed an SVM model to achieve attribution classification. However, this study only used homogeneous topological features, neglecting tactic and technique features, as well as text features. Moreover, this approach cannot fuse heterogeneous semantic information, limiting the comprehensiveness and accuracy of attribution analysis.

In summary, the current methods based on threat intelligence do not fully extract and fuse attribute type features, natural language text features, or topological relationship features. In addition, the importance of IOC attributes is ignored, and the heterogeneous information contained in CTI reports is not effectively organized and represented.

## 2.2 Cyber Threat Intelligence Modeling

Many models, knowledge bases, and analysis tools are currently related to threat intelligence[49]. STIX[50] is a structured threat information expression framework that can describe various types of information, such as IOCs, TTPs, and attack incidents, and support multiple use cases, including sharing threat information and analyzing network threats. The ATT&CK framework[51] describes network attack behavior from the perspectives of TTPs. It comprehensively and systematically summarizes the known knowledge of network attack techniques, expressing the relationships between tactics and techniques in a matrix. CVE[52] is an authoritative repository of vulnerability knowledge that assigns and registers numbers for publicly disclosed vulnerabilities and security issues concerning computer software and hardware. It allows different security groups to accurately describe and exchange information about vulnerabilities. Avclass2[53] is an open-source tool used for malware classification. It provides a hierarchical structure of malware classification labels, which are divided into four categories: behavior (BEH), category (CLASS), file attributes (FILE), and family (FAM). VirusTotal[54] is a tool that provides multiengine analyses of malware, IPs, domains, and other IOCs. It offers detailed IOC attribute information and associated data.

The works mentioned above have reviewed the structured representations of threat information from various perspectives, providing valuable insights for constructing the heterogeneous attributed graph schema for APT actor attribution in this paper.

## 2.3 Graph Representation Learning

Graph representation learning aims to embed graph data into a low-dimensional space while preserving the graph structure or attribute information. Several effective graph embedding methods, such as DeepWalk[55] and Node2vec[47], have been proposed to address homogeneous graph representation learning problems. Compared to those of extensively studied homogeneous graphs, the heterogeneity of heterogeneous graphs increases the difficulty of representation learning. Dong et al.[56] proposed the Metapath2vec method, which combines metapath-based random walks and a skip-gram model to learn heterogeneous graph embeddings. However, using only metapath-based random walks leads to the neglect of useful information such as node attributes. Graph neural networks (GNNs) have been proposed for extending deep neural networks to address graph data[57,58]. In particular, the graph attention networks (GATs)[58] introduce an attention mechanism to effectively learn the importance of nodes in a homogeneous graph. Based on this, Wang et al.[48] proposed the heterogeneous graph attention network (HAN), which utilizes node-level and semantic-level attention to learn the importance of nodes and metapaths for generating node embeddings in heterogeneous graphs. However, this method discards information of intermediate nodes along the metapath. In addition, Jin et al.[59] introduced a general framework for Heterogeneous Graph Neural Network via Attribute Completion (HGNN-AC), which performs end-to-end heterogeneous graph representation learning with attribute completion and reconstruction as auxiliary tasks.

This paper first introduces heterogeneous graph representation learning methods into APT actor attribution research, aiming to achieve more comprehensive and accurate attribution analysis by fusing heterogeneous graph information.

# 3. Method

## 3.1 Overall Framework

As the core of CTI, IOCs have numerous types, rich attributes, and various relationships. Therefore, we study APT actor attribution from the perspective of IOCs and characterize APT reports with their included IOCs through a heterogeneous attributed graph. We design a heterogeneous attributed graph schema for APT actor attribution based on various CTI models. To construct comprehensive IOC node representations, we extract and fuse the node features of three modalities: attribute types, natural language texts, and topological relationships. To further learn the deep hidden features of APT report nodes, we propose multilevel heterogeneous graph attention networks to weight and integrate the contributions provided by neighbors with different IOC types, different metapath-based report neighbors, and different metapaths.

The framework of our proposed APT-MMF method is shown in Figure 1. It consists of multimodal node feature extraction and multilevel report feature learning layers. In the multimodal node feature extraction layer, we extract three types of modal features based on the designed heterogeneous attributed graph schema. Attribute type features represent the categorical information of the node attribute values and are extracted via sequential or ID encoding. Natural language text features represent the contextual semantic information of node attribute values, and we use the BERT pretraining model and a fine-tuning layer to extract them. Topological relationship features represent the relation information of nodes and are extracted by the Node2vec algorithm. Through a feature-level fusion process implementing by concatenating the three feature modes, we obtain multimodal node features that can comprehensively represent node characteristics.

In the multilevel report feature learning layer, we introduce multilevel heterogeneous graph attention networks based on triple attention mechanisms to learn the deep hidden features of the report nodes. We first propose IOC type-level attention to fuse the information from neighbor nodes with different IOC types, thereby completing the sparse features of the attributeless report nodes derived from the multimodal extraction and fusion. Based on this level, we design twenty metapaths and integrate metapath-based neighbor node-level and metapath semantic-level attention[48] to further fuse multiple heterogeneous information. Specifically, the metapaths have reports as their end nodes and IOCs as their intermediate nodes; metapath-based neighbor node-level attention fuses the completed features of different report neighbor nodes based on one metapath; metapath semantic-level attention further fuses the previously learned report node features based on different metapaths. We input the final learned deep hidden features of the report nodes into a fully connected layer for classification, obtaining attribution results.

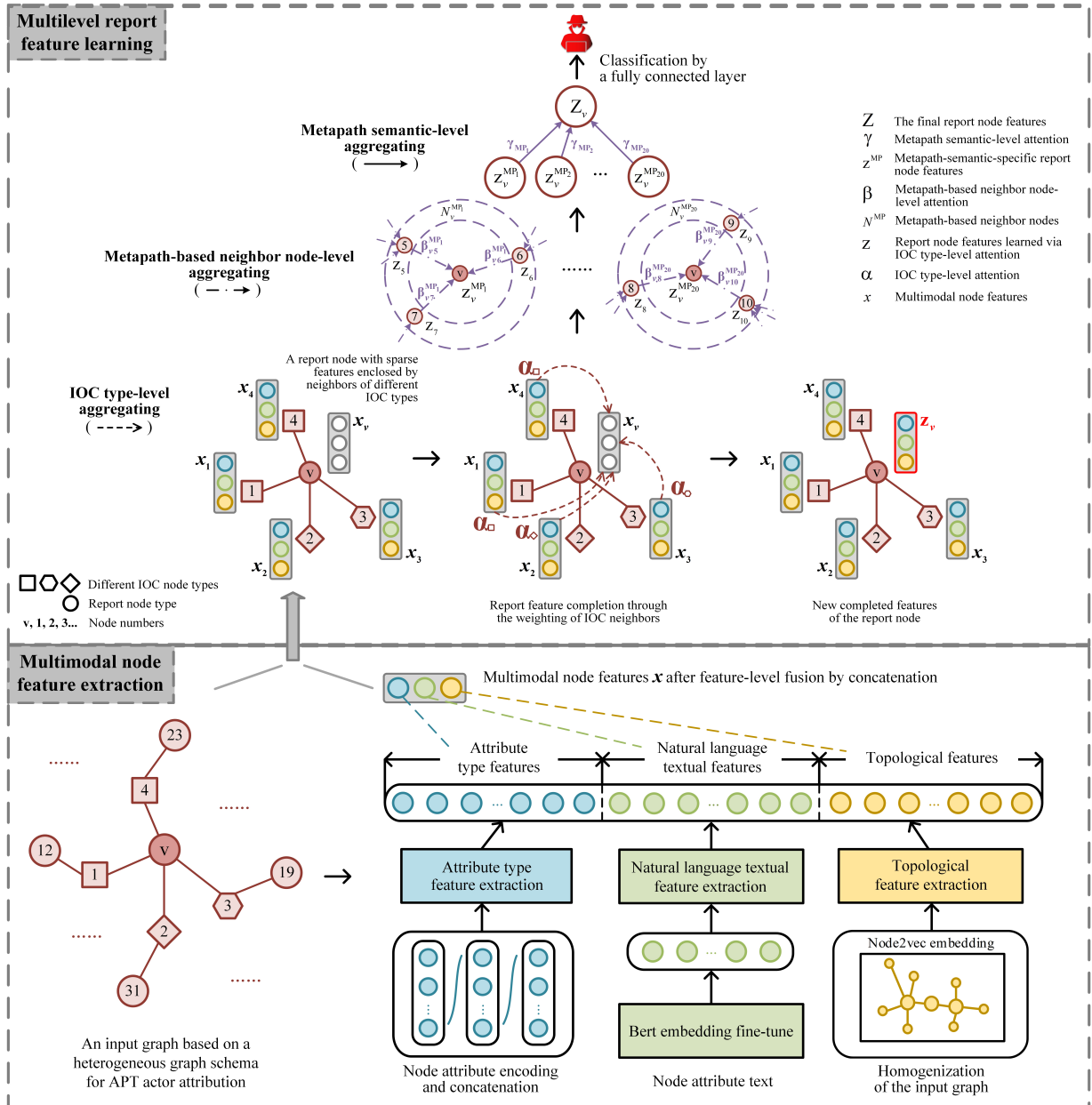

**Figure 1. The framework of APT-MMF**

## 3.2 Heterogeneous Attributed Graph Schema for APT Actor Attribution

The heterogeneous attributed graph schema for APT actor attribution characterizes APT reports and their IOC information. It specifies node types, node attribute types, and node relationship types. This approach is an important component of our method and is the basis for constructing the utilized graph dataset. We use the STIX format[50], integrate ATT&CK[51] and CVE[52] knowledge bases, and expand the information provided by Avclass2[53] and Virustotal[54] to design the schema. Based on these CTI models, we select the crucial IOC types and their attribute types as node types and node attribute types in the schema, respectively. We also connect the IOC types with the APT report node type, which are the centers of the schema. In addition, we select the crucial IOC relations based on these CTI models and incorporate them into the node relationship types of the schema. The designed heterogeneous attributed graph schema for APT actor attribution is illustrated in Figure 2.

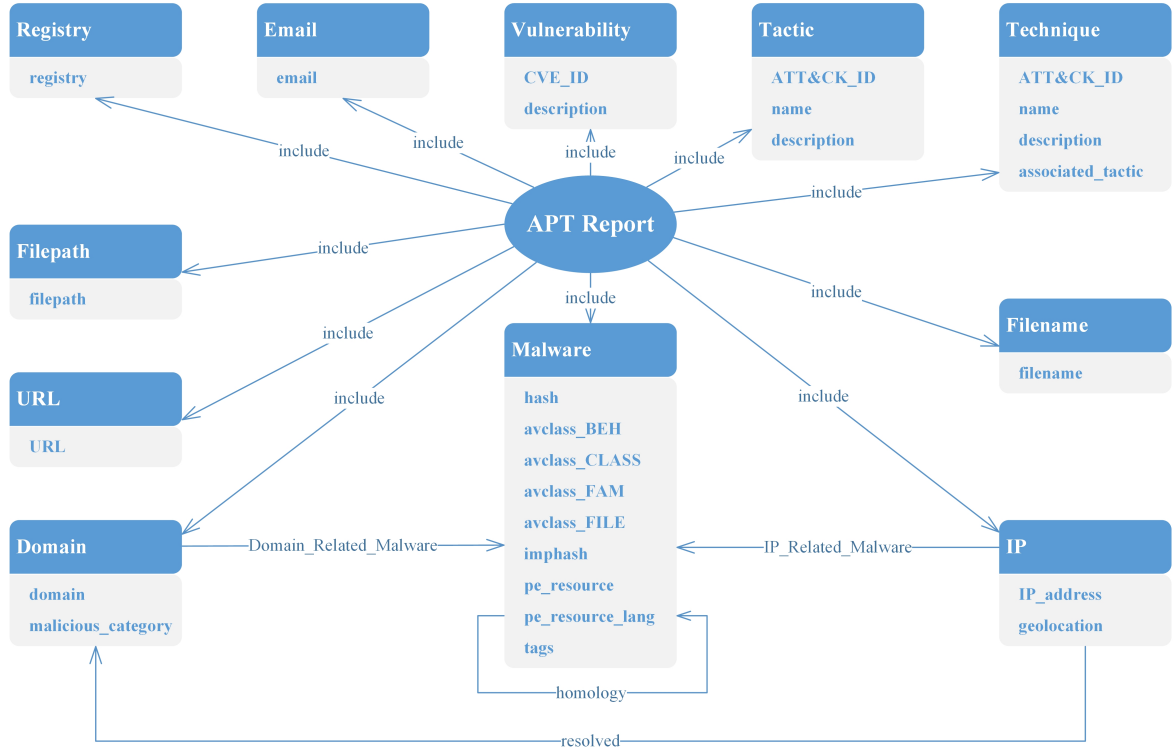

Figure 2. Heterogeneous attributed graph schema for APT actor attribution

We select eleven IOC types based on the IOC types contained in STIX [50] and their common occurrences in APT reports, including malware, tactics, techniques, vulnerabilities, IPs, domains, URLs, filenames, file paths, registries, and emails. These IOC types are incorporated as node types into the heterogeneous attributed graph schema. We add attribute types based on specific CTI knowledge base schemas or result fields acquired from commonly used analysis tools for the important IOC types listed below.

1. Malware: This is an arsenal of APT groups characterized by a unique hash value. We supplement the attribute types based on the analysis result fields obtained from Avclass[53] and VirusTotal[54], such as "avclass_BEH", "imphash", and "pe-resource". The attribute names and their descriptions are listed in Table 1.
2. Tactics and techniques: Based on the knowledge model of ATT&CK[51], we add three attribute types, namely, "ATT&CK_ID", "name" and "description". In addition, we add the attribute type "associated_tactic" to the technique node types.
3. Vulnerabilities: Based on the CVE knowledge base model[52], we add two attribute types, namely "CVE_ID" and "description".
4. IPs: Based on the analysis result fields obtained from VirusTotal[54], we add the attribute type

"geolocation" to the IP node.

5. Domains: Based on the analysis result fields acquired from VirusTotal[54], we add the attribute type "malicious category", whose values are similar to "phishing and fraud" and "command and control".

The node relationship types mainly contain five categories: "Inclusion" between a report type and all other types; "Resolution" between the IP type and the domain type; "Association between IP type and Malware type", specifically indicating that malware includes IP strings, communicates with an IP, or downloads information from an IP address; "Association between domain type and malware type", specifically indicating that malware includes domain strings, communicates with a domain, or downloads information from a domain; and "Homology" between malware types.

Table 1. The attributes of malware

| Attribute Name | Description |
| --- | --- |
| hash | Malware hash value |
| avclass_BEH | Malware behavior, such as information theft and mining |
| avclass_CLASS | Malware categories, such as backdoor and ransomware |
| avclass_FAM | Malware family |
| avclass_FILE | Static file properties of malware, including the file type, the operating system used to execute the sample, the packer used for obfuscation, and the programming language used to encode the sample |
| imphash | Hash values of the imported functions contained in malware |
| pe-resource | Resources used by malware, such as the strings, icons, and data packaged by the malware |
| pe-recource_lang | The programming language employed for crafting the resources utilized by malware |
| tags | The collection of malware tags |

An illustrative example of our heterogeneous attributed graph schema for APT actor attribution is shown in Figure 3. The graph includes four APT report nodes connected through the nodes of various IOC types, such as malware, URLs, domains, filepaths, registries, and techniques.

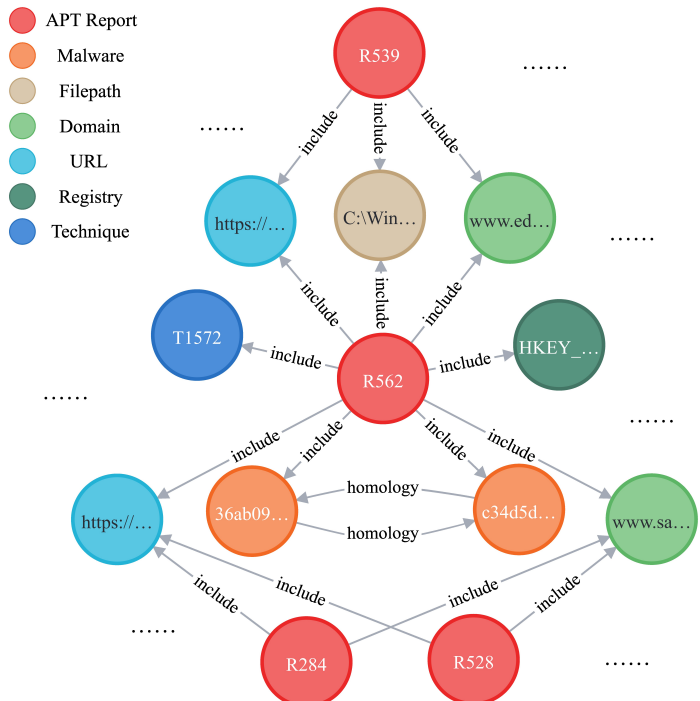

Figure 3. An example of a heterogeneous attributed graph for APT actor attribution

## 3.3 Multimodal Node Feature Extraction and Fusion

To comprehensively represent node characteristics, we extract features from three modalities, namely, attribute types, natural language text, and topological relationships, and then fuse them. The attribute type features represent the categorical information of the node attribute values. The natural language text features represent the contextual semantic information of the node attribute values. The topological relationship features represent the relation information of the nodes. We conduct feature-level fusion by concatenating these three types of modal features and obtaining a multidimensional feature vector for each node. We summarize the designs of the multimodal node features with their names, descriptions, encoding methods, and dimensions in Table 2.

Table 2. The designs of multimodal node features

| Feature name | | Description | Encoding | Dimensions |
|---|---|---|---|---|
| Attribute Type Feature | Node type | Type of node | Ordinal encoding | 2 |
| | Domain: malicious category | Malicious domain category | Ordinal encoding | 2 |
| | IP: geolocation | Geolocation of an IP | Ordinal encoding | 2 |
| | IP: ip_address | Address of an IP | ID encoding | 12 |
| | Tactic & technique: ATT&CK_ID | ATT&CK_ID of a tactic and technique | ID encoding | 11 |
| | Vulnerability: CVE_ID | CVE_ID of a vulnerability | ID encoding | 9 |
| | Malware: hash | Malware hash value | Ordinal encoding | 4 |
| | Malware: avclass_BEH | Malware behavior, such as information theft and mining | Ordinal encoding | 2 |
| | Malware: avclass_CLASS | Malware categories, such as backdoor and ransomware | Ordinal encoding | 2 |
| | Malware: avclass_FAM | Malware family | Ordinal encoding | 2 |
| | Malware: avclass_FILE | Static file properties of malware, including the file type, the operating system used to execute the sample, the packer used for obfuscation, and the programming language used to encode the sample | Ordinal encoding | 2 |
| | Malware: imphash | Hash values of the imported functions contained in malware | Ordinal encoding | 4 |
| | Malware: pe_resource | Resources used by malware, such as the strings, icons, and data packaged by the malware | Ordinal encoding | 4 |
| | Malware: pe_resource_lang | The programming language employed for crafting the resources utilized by malware | Ordinal encoding | 2 |
| | Malware: tags | The collection of malware tags | Ordinal encoding | 4 |
| Natural language text feature | | Contextual semantic information of node attribute values | BERT encoding | 64 |
| Topological relationship feature | | Relation information of nodes | Node2vec encoding | 128 |

We specify the processes used to extract attribute type features, natural language text features, and topological relationship features below.

**(1) Attribute type features**

Some attribute types have categorical attribute values. The categorical information of node attribute values is an important node information modality. Based on the heterogeneous attributed graph schema, we select fifteen relevant attribute types and design 64-dimensional attribute type features, as shown in Table 2. These fifteen types include node types, malicious domain categories, IP geolocations, all attribute types of malware, and ID-like attribute types (including IP addresses, CVE IDs of vulnerabilities, and ATT&CK IDs of tactics and techniques). The values of ID-like attribute types often contain sets of numbers with fixed formats and are distinguished from those of the other attribute types. Therefore, we adopt different feature encoding approaches for ID-like and non-ID-like attribute types.

We design the specific feature encoding method for ID-like attribute types based on their format. For tactics and techniques, ATT&CK IDs utilize formats such as abbreviated letters, fixed-length numbers, and dot separators; examples include the tactic "TA0011", the technique "T1071", and the sub-technique "T1071.004". Accordingly, we design 11-dimensional features. The first four digits are the numbers in the ATT&CK ID of the tactic, the middle four digits are the numbers in the ATT&CK ID of the technique, and the last three digits are the last three digits in the ATT&CK ID of the sub-technique. For example, the feature of the tactic "TA0011" is $(00110000000)^T$, the feature of the technique "T1071" under tactic "TA0011" is $(00111071000)^T$, and the feature of the sub-technique "T1071.004" under technique "T1071" is $(00111071004)^T$. For an IP, specifically IPv4, which is common in CTI reports, the address consists of four groups of numbers separated by dots, and each group of numbers contains a maximum of three digits. Accordingly, we design 12-dimensional features concatenated with these four groups of numbers, which are padded with zeros from the front if the length of the numbers in the group is less than 3. For example, the feature of the IP address "45.135.167.27" is $(045135167027)^T$. For a vulnerability, the CVE ID is identified using abbreviated letters and two groups of numbers separated by a hyphen. The first group of numbers contains four digits, and the second group contains a maximum of five digits. Accordingly, we design 9-dimensional features concatenated with the two groups of numbers, which are padded with zeros from the front if the length of the numbers in the second group is less than 5. For example, the feature of the vulnerability "CVE-2019-9670" is $(201909670)^T$.

Non-ID-like attribute types involve many categorical values. Performing one-hot encoding on these attributes would result in sparse features and a computational loss. Therefore, we adopt ordinal encoding rather than one-hot encoding. Specifically, for each attribute type, we build a vocabulary sorted in character order of the values and utilize a fixed-length numeric index as the feature of the corresponding value. The feature dimensionality of the attribute type is equal to the digits of the corresponding vocabulary length.

Since these fifteen types are independent, we concatenate their features to construct 64-dimensional attribute type features for each node. The concatenation order is the same as that shown in Table 2. In particular, for the missing value of an attribute type, we utilize the zero vector of the corresponding dimension as the feature.

**(2) Natural language text features**

The values of some attribute types have prominent natural language characteristics, such as the descriptions of tactics and the names of domains. The contextual semantic information of these text values is a node information modality, and the associated semantic similarity contributes to threat actor attribution. Therefore, we extract natural language text features as a component of the multimodal node features. We use BERT to encode these attribute values so that the text features of semantically similar attribute values are close together. BERT[1] is a natural language model with excellent semantic representation capabilities, and it can be efficiently fine-tuned for different downstream tasks[46]. Table 3 lists the node and attribute types related to the natural language text features in the heterogeneous attributed graph schema. For each node, we input the corresponding attribute values into the pretrained BERT model and take the mean value of the last hidden state as the output features. Then, we input these features into a fully connected layer for dimensionality reduction and fine-tuning, obtaining 64-dimensional natural language text features for each node. In particular, we input the empty string for missing values.

Table 3. Node and attribute types related to natural language text features

| Node type | Attribute type |
| --- | --- |
| Malware | avclass_BEH, avclass_CLASS, avclass_FAM, avclass_FILE, pe_resource_lang, tags |
| Tactic | name, description |
| Technique | name, description |
| Vulnerability | description |
| Domain | name, malicious_category |
| IP | ip_address |
| Registry | registry |
| Filename | filename |
| Filepath | filepath |
| Email | email |
| URL | url |

**(3) Topological relationship features**

Topological relationships form another important node information modality. Extracting topological relationship features and integrating them into multimodal node features is a way to explicitly utilize topological information, and this strategy can complement GNNs to enhance the use of topological information and achieve improved performance. We extract topological relationship features through Node2vec[47]. Node2vec is a continuous node feature representation learning algorithm based on biased random walks for homogeneous graphs. We ignore heterogeneous information such as the node types and attributes encoded into attribute type features or natural language text features. We thereby input the graph data into Node2vec as a homogeneous graph and obtain 128-dimensional topological relationship features for each node.

## 3.4 Multilevel Heterogeneous Graph Attention Networks for Deep Feature Learning of Report Nodes

To further learn the deep hidden features of the report nodes, we propose multilevel heterogeneous graph attention networks for fusing different heterogeneous graph information with different importance. Briefly, IOC type-level attention fuses node information acquired from the IOC neighbors of the report nodes; metapath-based neighbor node-level attention and metapath semantic-level attention fuse the node information of metapath-based report neighbors and metapath semantic information, respectively. Overall, IOC type-level attention fuses the information of intermediate nodes along the metapaths discarded by the metapath-based neighbor node-level and metapath semantic-level attention mechanisms. These triple attention mechanisms progress in a hierarchical manner and complement each other.

---

[1] https://huggingface.co/bert-base-cased

**(1) Report node feature completion based on IOC type-level attention**

The report node features derived from multimodal extraction and fusion are sparse due to the attributeless nature of report nodes, especially for attribute type features and natural language text features. The IOC information contained in an APT report is scattered among the first-order IOC neighbors of the report node. More contextual IOC information is contained in the second-order IOC neighbors of the report node. We thereby utilize the first-order and second-order IOC neighbors of the report node to complete the sparse report features. We observe that different types of IOC neighbors provide different contributions to APT actor attribution. Some IOC types, such as malware and domains, are usually more crucial for APT actor attribution than others, such as filenames. Therefore, we propose an IOC type-aware attention mechanism for learning the importance of different IOC neighbor nodes and aggregate their information to complete the report node features.

Suppose we are given a node pair $(v, u)$, where $v$ is the report node and $u$ is the first-order or second-order IOC neighbor node of $v$. Let the set of first-order and second-order IOC neighbor nodes of report node $v$ be denoted as $N_v$; then, $u \in N_v$. The attention layer can learn the associated importance coefficient $e_{vu}$, that is, the contribution of node $u$ to node $v$. The importance coefficient $e_{vu}$ is related to the IOC type of node $u$, that is:

$$e_{vu} = \sigma(W h_u) \tag{1}$$

where $\sigma$ is the $LeakyReLU(\cdot)$ activation function, $W \in \mathbb{R}^{d \times 1}$ is the attention parameter matrix ($d$ is the number of IOC types), and $h_u$ is the one-hot vector for the IOC type of node $u$.

After obtaining the importance coefficients of all neighbors, we calculate the normalized weighted coefficient $a_{vu}$ with the $softmax(\cdot)$ function:

$$a_{vu} = softmax(LeakyReLU(e_{vu})) = \frac{\exp(LeakyReLU(e_{vu}))}{\sum_{s \in N_v} \exp(LeakyReLU(e_{vs}))} \tag{2}$$

Let the multimodal node features be denoted as $x$. We aggregate the features of neighbor nodes according to the normalized weighted coefficient $a_{vu}$ to obtain the completed feature $z_v^C$ of the report node $v$:

$$z_v^C = \sum_{u \in N_v} a_{vu} x_u \tag{3}$$

The above attention process is extended to multihead attention to stabilize the learning process and reduce the high variance caused by graph heterogeneity. In this way, the completed feature of report node $v$ can be rewritten as follows:

$$z_v^C = mean\left(\sum_k^K \sum_{u \in N_v} a_{vu} x_u\right) \tag{4}$$

where $K$ indicates that we perform $K$ independent attention processes and $mean(\cdot)$ denotes that we average the $K$ results.

Finally, we use the completed features to update the original features of the report node $v$ and obtain the report features $z_v$ learned by the IOC type-level attention mechanism:

$$z_v^C = mean\left(\sum_k^K \sum_{u \in N_v} a_{vu} x_u\right) \tag{5}$$

**(2) Heterogeneous feature fusion according to metapath-based neighbor node-level and metapath semantic-level attention**

In addition to the heterogeneous information derived from IOC types, various pieces of heterogeneous graph information can be extracted from multiple predesigned metapaths. We consider three perspectives for designing metapaths based on the heterogeneous attributed graph schema developed for APT actor attribution:

- Node and relationship types: The heterogeneous attributed graph schema contains various node and relationship types. A set of metapaths must comprehensively cover this heterogeneous information.
- Metapath length and complexity: The accuracy and efficiency of analytic calculations can be affected by the lengths and complexity levels of metapaths. Metapaths that are excessively long result in sparse metapath-based neighbors and degrade the performance of the utilized model.
- Metapath diversity: Different metapaths contain different semantic information, and various metapaths can express richer and more comprehensive information.

We denote the number of nodes traversed from the head node to the tail node as the order of the examined metapath. Based on the above considerations, we determine the scopes of metapaths as first-order to fourth-order metapaths, with the report nodes as the end nodes and the IOC nodes as the intermediate nodes. We further design twenty such metapaths according to our heterogeneous attributed graph schema, which is distinguishable for attribution, as shown in Table 4. These metapaths characterize heterogeneous semantic information from multiple perspectives, which is beneficial for obtaining more comprehensive report node representations.

**Table 4. The metapaths of our method**

| Order | Number | Metapath | Description |
|---|---|---|---|
| First-order | $MP_1$ | report-filename-report | Two reports include the same filename |
| | $MP_2$ | report-malware-report | Two reports include the same malware |
| | $MP_3$ | report-URL-report | Two reports include the same URL |
| | $MP_4$ | report-filepath-report | Two reports include the same filepath |
| | $MP_5$ | report-domain-report | Two reports include the same domain |
| | $MP_6$ | report-registry-report | Two reports include the same registry |
| | $MP_7$ | report-IP-report | Two reports include the same IP |
| | $MP_8$ | report-vulnerability-report | Two reports include the same vulnerability |
| | $MP_9$ | report-tactic-report | Two reports include the same tactic |
| | $MP_{10}$ | report-email-report | Two reports include the same email |
| Second-order | $MP_{11}$ | report-IP-malware-report | The IP included in Report A and the malware included in Report B are related |
| | $MP_{12}$ | report-domain-malware-report | The domain included in Report A and the malware included in Report B are related |
| | $MP_{13}$ | report-malware-malware-report | The malware included in Report A and the malware included in Report B have the same origin |
| Third-order | $MP_{14}$ | report-IP-malware-IP-report | The IP included in Report A and the IP included in Report B are related to the same malware |
| | $MP_{15}$ | report-IP-domain-IP-report | The IP included in Report A and the IP included in Report B can be resolved from the same domain |
| | $MP_{16}$ | report-domain-IP-domain-report | The domain included in Report A and the domain included in Report B can be resolved to the same IP |
| | $MP_{17}$ | report-malware-domain-malware-report | The malware included in Report A and the malware included in Report B are related to |

| | | | the same domain |
|---|---|---|---|
| | MP$_{18}$ | report-malware-IP-malware-report | The malware included in Report A and the malware included in Report B are related to the same IP |
| Fourth-order | MP$_{19}$ | report-IP-malware-malware-IP-report | The malware related to the IP in Report A and the malware related to the IP in Report B have the same origin |
| | MP$_{20}$ | report-domain-malware-malware-domain-report | The malware related to the domain in Report A and the malware related to the domain in Report B have the same origin |

To fuse metapath-based heterogeneous graph information, we integrate the metapath-based neighbor node-level attention and metapath semantic-level attention[48] to update the features of each report node. Metapath-based neighbor node level attention learns the importance of different report neighbors of a report node based on one metapath. It weights and aggregates the information of the report neighbors while discarding the information of intermediate nodes along the metapath that are IOC types. Metapath semantic-level attention further learns the importance of different metapaths and performs weighted fusion on the report node features learned through metapath-based neighbor node-level attention.

We input the report features that are finally learned through the triple attention mechanisms into a fully connected layer to obtain the attribution classification results. The entire model training process is optimized by minimizing the cross-entropy loss function between the ground-truth labels and the predicted labels:

$$L = -\sum_{l \in \mathcal{Y}_L} Y^l \ln(C \cdot Z^l) \qquad (6)$$

where $C$ represents the parameters of the classifier, $\mathcal{Y}_L$ represents the index set of the report nodes in the training set, and $Y^l$ and $Z^l$ represent the ground-truth labels and deep features of the report nodes, respectively.

## 4. Experiments

### 4.1 Datasets

Due to the lack of public graph datasets with IOCs for APT actor attribution, we use multisource threat intelligence to construct a heterogeneous attributed graph dataset for experiments. The data sources include the APTNotes repository[2]; threat intelligence shared by security vendors such as Symantec[3], FireEye[4], Kaspersky[5], Qianxin[6], and Nsfocus[7]; the knowledge bases of CVE[8] and ATT&CK[9]; and the result data produced by common security analysis tools such as VirusTotal and Avclass2. Among them,

---

[2] https://github.com/aptnotes/data/

[3] https://symantec-enterprise-blogs.security.com/blogs/threat-intelligence

[4] https://www.trellix.com/en-us/advanced-research-center/threat-intelligence.html

[5] https://www.kaspersky.com/enterprise-security/threat-intelligence

[6] https://ti.qianxin.com/blog/

[7] https://nti.nsfocus.com/

[8] https://cve.mitre.org/

[9] https://attack.mitre.org/

APTNotes and threat intelligence from various vendors are the primary sources of APT reports, and the other data sources are mostly used to supplement the relations and attributes of IOCs.

The process of constructing the dataset is divided into multiple steps. We first collect and select APT reports from the above data sources. We prioritize the reports released after 2015. Through manual inspection, we screen out reports that have clear and unique APT actor attribution and contain rich IOC information. We ultimately collect 1,300 reports belonging to 21 APT groups, as shown in Figure 4. These collected APT reports are unstructured texts, which must be further processed through entity extraction, entity cleaning, and entity attribute and relationship expansion.

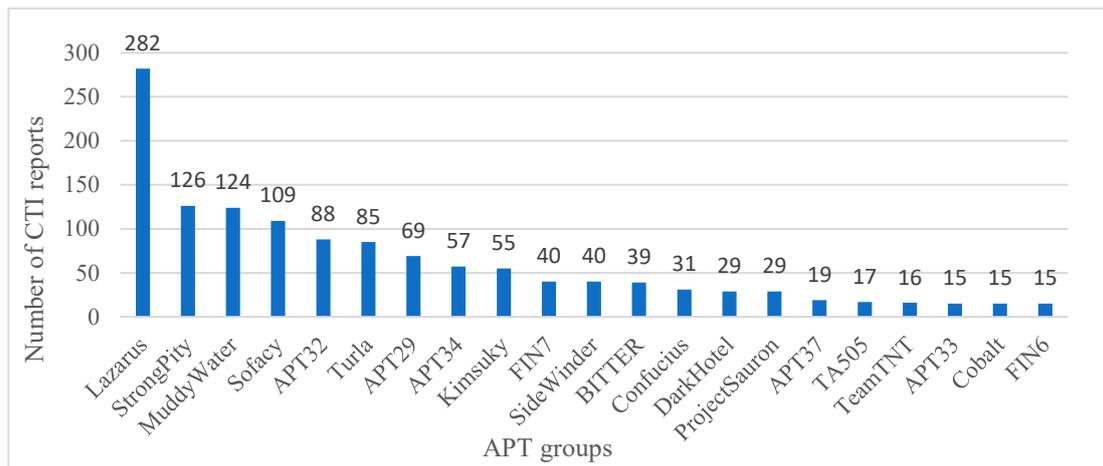

**Figure 4. Number of CTI reports associated with each APT group**

**Entity extraction.** The tactics and techniques in APT reports are usually presented in various textual descriptions, while other IOC entities, such as IPs and domains, have textual formats. Therefore, we use different extraction approaches for different entities. For tactics and techniques, we use TRAM[10] to map unstructured APT reports to the ATT&CK framework. TRAM vectorizes sentences using a bag-of-words model and employs machine learning algorithms to classify them according to their tactics and techniques while providing confidence in the classification results. We retain only the results with 100% confidence as the tactic and technique extraction results for APT reports. For other IOC entities, such as IPs and domains, we use IOC-Finder[11] to extract them from the APT reports. We improve the regular expressions of IOC-Finder to compensate for shortcomings such as false positives in IP extraction, false negatives in filepath extraction, and incomplete registry extraction, thereby optimizing the extraction effect of these IOC entities.

**Entity cleaning.** We find that the extracted IOCs are mixed with some nonmalicious entities, and this issue primarily occurs for three types: IPs, domains, and URLs. Therefore, it is necessary to perform entity cleaning specifically for these three types. During the cleaning process, we use methods such as whitelist filtering, VirusTotal static analysis, and identification of invalidation handling. For IPs, we first use a whitelist to filter out the nonmalicious entities. Then, we utilize the VirusTotal static analysis process to detect whether each of the filtered IPs is associated with malicious URLs. If corresponding malicious URLs are observed, which means that the IP causes malicious behaviors, the IP is judged as malicious. Otherwise, the IP is filtered out. For domains, the cleaning process is the same as that applied for IPs. For URLs, since the invalidation handling, such as [.] or hxxp, is often used for malicious URLs to prevent accidental clicks from causing harm, we first identify whether each URL contains the invalidation handling. If there is the invalidation handling, the URL is judged as malicious. Otherwise, it is judged through the VirusTotal static analysis. If the number of engines in the analysis results that judge a URL as malicious is greater than 10, the URL is judged as malicious. Otherwise, the URL is filtered out. Afterward, we establish inclusion relationships between the reports and the extracted and cleaned IOCs.

---

[10] https://github.com/center-for-threat-informed-defense/tram/

[11] https://github.com/fhightower/ioc-finder/

**Entity attribute and relationship expansion.** We expand the attributes and relationships of different IOC entities based on various data sources. According to the ATT&CK knowledge base, we expand the name and description attributes of both tactics and techniques, as well as the associated tactic attribute of the techniques. With the CVE knowledge base, we expand the description attributes of the vulnerabilities. Based on the Avclass2 analysis results, we expand the four Avclass attributes of malware and establish a homologous relationship among the malware whose Avclass_FAM attributes are matched. Based on the VirusTotal static analysis results, we expand the geolocation attributes of IPs, the malicious category attributes of domains, and the other malware attributes except for the hash and Avclass attributes. We also expand the IP, domain, and malware entities based on the VirusTotal static analysis results and establish resolution, association, and homologous relationships among these types of entities.

The constructed heterogeneous attributed graph dataset contains 24694 nodes and 40335 relationships. The number of nodes extracted directly from APT reports is 15540. The remaining nodes are expanded from VirusTotal. The number of nodes and relationships belonging to various types are shown in Figure 5 and Figure 6, respectively. We divide the report nodes under each APT group label at a ratio of 8:1:1 to form the training, validation, and test sets.

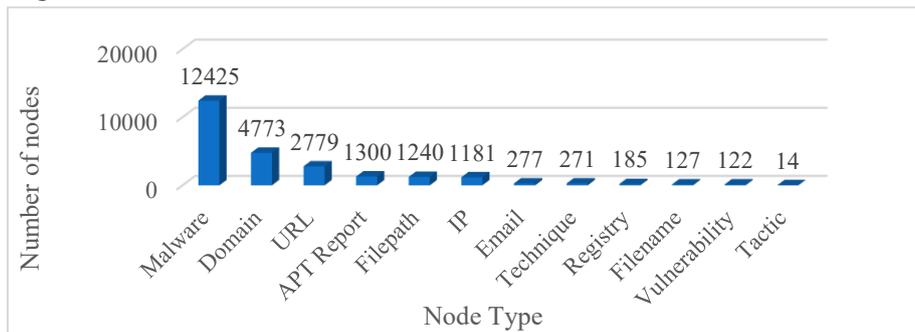

**Figure 5. Number of nodes belonging to various types**

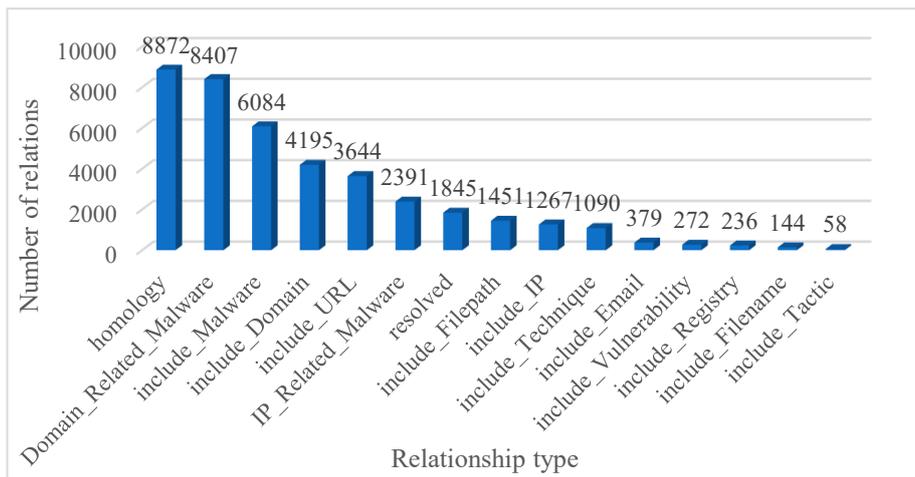

**Figure 6. Number of relationships belonging to various types**

## 4.2 Experimental Settings

To quantitatively evaluate the proposed method, we use the $Micro\text{-}F_1$ and $Macro\text{-}F_1$ metrics, which are common evaluation indicators in multiclassification tasks[60]. $Micro$-$F_1$ calculates the overall precision and recall of all the labels. Let $TP_t$, $FP_t$, and $FN_t$ denote the true positives, false positives, and false negatives, respectively, of the $t$th label in the label set $L$. $Micro$-$F_1$ can be defined as follows:

$$Micro - F_1 = \frac{2 \times Precision \times Recall}{Precision + Recall} \tag{7}$$

$$Precision = \frac{\sum_{t \in L} TP_t}{\sum_{t \in L} TP_t + \sum_{t \in L} FP_t} \tag{8}$$

$$Recall = \frac{\sum_{t \in L} TP_t}{\sum_{t \in L} TP_t + \sum_{t \in L} FN_t} \tag{9}$$

$Macro$-$F_1$ is another kind of $F_1$ score that calculates the average $F_1$ score under each label. $Macro$-$F_1$ can be defined as follows:

$$Macro - F_1 = \frac{1}{|L|} \sum_{t \in L} \frac{2 \times Precision_t \times Recall_t}{Precision_t + Recall_t} \tag{10}$$

$$Precision_t = \frac{TP_t}{TP_t + FP_t} \tag{11}$$

$$Recall_t = \frac{TP_t}{TP_t + FN_t} \tag{12}$$

We conduct four groups of experiments:

**Experiment 1**: Comparison experiment. We compare our model with seven machine learning models from related studies[35–41], such as the Random Forest, XGBoost, and MLP models. In addition, since the GNN is an important foundation of our method, we compare our method with four GNN models such as GAT[58] and HAN[48].

**Experiment 2**: Effectiveness experiment for multimodal node features. The multimodal node features in our method contain multiple dimensions. Taking the malware part of attribute type features as the base, we successively add the other part of attribute type features, natural language text features, and topological relationship features to prove the effectiveness of each set of features.

**Experiment 3**: Effectiveness experiment for triple attention aggregation. The multilevel heterogeneous graph attention networks in our method contain the triple attention mechanisms. Taking the metapath-based neighbor node-level attention as the base, we successively add the metapath semantic-level attention after it and our proposed IOC type-level attention before it to prove the effectiveness of each level of attention aggregation.

**Experiment 4**: Effectiveness experiment for metapaths. We design twenty metapaths for metapath-based neighbor node-level and metapath semantic-level attention in the triple attention mechanisms. Taking the first-order metapaths as the base, we successively add the second-order metapaths, the third-order metapaths, and the fourth-order metapaths to prove the effectiveness of each set of metapaths.

## 4.3 Results

### 4.3.1 Comparison Experiment

As mentioned in the related work section, the main idea of the existing APT actor attribution methods is to classify APT report features based on traditional machine learning models[35–41]. We compare the models used in these studies with our method. These models include Naïve Bayes[35], KNN[35], Decision Tree[35,38], SVM[38,41], Random Forest[35,38], XGBoost[39], and MLP[40]. For each report, we sum the multimodal features of the corresponding report node and those of the IOC nodes that are directly included in the report. We input the resulting sum as the APT report features into the above comparison models.

We also select four GNN models as comparison methods. We input the adjacency matrix and multimodal node features into these comparison models. The graph convolutional network (GCN)[57] is a homogeneous graph neural network that uses an efficient layer-wise propagation rule based on the first-order approximation of spectral graph convolutions on graphs. We evaluate the GCN on multiple metapath-based homogeneous graphs and report the best results. The GAT[58] is a homogeneous graph neural network that uses masked self-attention layers to process graph-structured data. We evaluate the GAT on multiple metapath-based homogeneous graphs and report the best results. The HAN[48] is a heterogeneous graph neural network that utilizes node-level and semantic-level attention to learn node embeddings. We use all the metapaths designed in this paper for the HAN calculations. The HGNN-AC[59] is an advanced heterogeneous graph neural network framework that proposes an attention mechanism based on topological node embeddings for node attribute completion and can be combined with any heterogeneous graph neural networks for end-to-end learning. We use Metapath2vec[56] to compute topological node embeddings and combine HGNN-AC with HAN[48]. The results of each comparison method are shown in Table 5.

Table 5. Results of comparison experiments

| Category | Methods | Micro-$F_1$ | Macro-$F_1$ |
|---|---|---|---|
| Traditional ML Models | Naïve Bayes | 0.4379 | 0.3972 |
| | KNN | 0.4598 | 0.3367 |
| | Decision Tree | 0.4744 | 0.2707 |
| | SVM | 0.5401 | 0.3825 |
| | Random Forest | 0.6788 | 0.5540 |
| | XGBoost | 0.7372 | 0.5929 |
| | MLP | 0.7445 | 0.4869 |
| GNNs | GCN | 0.7518 | 0.5693 |
| | GAT | 0.7737 | 0.6641 |
| | HAN | 0.7810 | 0.6838 |
| | HGNN-AC | 0.8029 | 0.6871 |
| | Ours | **0.8321** | **0.7051** |

From the experimental results, our method achieves the best attribution effect and is significantly better than the comparison methods overall. Among the seven machine learning-based comparison models, the best approach is the MLP method[40]. Compared with this method, our method improves the $Micro\text{-}F_1$ indicator by 8.76% and the $Macro\text{-}F_1$ indicator by 11.82%. These results show that our method can more precisely characterize the heterogeneous IOC information contained in APT reports using heterogeneous attributed graph schema and metapaths. Moreover, by using the triple attention mechanisms, our method can effectively fuse various features and learn report features with richer and more accurate semantics.

Among the four GNN comparison methods, the homogeneous graph neural networks GCN and GAT cannot utilize heterogeneous graph information, which results in poor performance. HAN fuses the information of metapath-based neighbor nodes that belong to the report type but discards the information of intermediate nodes along metapaths that belong to the IOC type, which results in trivial performance. HGNN-AC adds node attribute completion based on topological attention, improving the performance compared to that of HAN. The topology of an IOC node may be more similar to the topology of a particular report node than those of other IOC nodes. However, the features of this IOC node are not necessarily more crucial to the attribution of the report than those of other IOC nodes. From this perspective, the completion strategy of HGNN-AC still exhibits inadequacy. Our method retains the node-level and semantic-level attention of HAN to fuse the heterogeneous information of metapath-based neighbors effectively. Moreover, we propose a completion strategy based on IOC-type attention, which is significantly better than that based on topological attention in HGNN-AC.

4.3.2 Effectiveness Experiment for Multimodal Node Features

The multimodal node features designed in our method contain many dimensions, as shown in Table 2. We divide these features into four sets, which are the Malware part of Attribute Type features (marked as "MAT"), the Other part of Attribute Type features (marked as "OAT"), Natural Language Text features (marked as "NLT"), and Topological Relationship features (marked as "TR"). Taking the malware part of attribute type features as the base, we successively add other sets of features and observe changes in evaluation indicators. The experimental results are shown in Table 6. The performance significantly improves each time a set of features is added, which proves the effectiveness of our designed multimodal node features.

Table 6. Effectiveness experimental results of multimodal node features

| Node Features | Micro-$F_1$ | Macro-$F_1$ |
| --- | --- | --- |
| MAT | 0.4672 | 0.2775 |
| MAT + OAT | 0.5912 | 0.4105 |
| MAT + OAT + NLT | 0.7518 | 0.6189 |
| MAT + OAT + NLT + TR | **0.8321** | **0.7051** |

4.3.3 Effectiveness Experiment for Triple Attention Aggregation

The multilevel heterogeneous graph attention networks contain the triple attention mechanisms. Our proposed IOC type-level attention plays a pivotal role in the entire method. It not only completes the sparse features of the attributeless report nodes derived from the prior multimodal extraction and fusion, but also fuses the IOC neighbor node information discarded by the subsequent two-level attention. We start the experiment with the metapath-based neighbor node-level attention and assign equal importance to each metapath. We then add the metapath semantic-level attention after it and finally introduce our proposed IOC type-level attention before it. The experimental results are shown in Table 7. The performance significantly improves each time a level of attention aggregation is added, which proves the effectiveness of triple attention aggregation.

Table 7. Effectiveness experimental results of triple attention aggregation

| Attention | Micro-$F_1$ | Macro-$F_1$ |
| --- | --- | --- |
| Metapath-based neighbor node level | 0.7445 | 0.5868 |
| Metapath-based neighbor node level + Metapath semantic level | 0.7810 | 0.6838 |
| Metapath-based neighbor node level + Metapath semantic level + IOC type level | **0.8321** | **0.7051** |

4.3.4 Effectiveness Experiment for Metapaths

Twenty metapaths are designed in our method and used for metapath-based neighbor node-level and metapath semantic-level attention in the triple attention mechanisms. We divide these metapaths into four sets, as shown in Table 4, which are ten first-order metapaths, three second-order metapaths, five third-order metapaths, and two fourth-order metapaths. Taking the first-order metapaths as the base, we successively add other sets of metapaths and observe changes in evaluation indicators. The experimental results are shown in Table 8. The performance significantly improves each time a set of metapaths is added, which proves the effectiveness of our designed metapaths.

Table 8. Effectiveness experimental results of metapaths

| Metapaths | Micro-$F_1$ | Macro-$F_1$ |
|---|---|---|
| First order | 0.7956 | 0.6919 |
| First order+Second order | 0.8029 | 0.6928 |
| First order+Second order+Third order | 0.8102 | 0.6997 |
| First order+Second order+Third order+Fourth order | **0.8321** | **0.7051** |

## 4.4 Explanatory Analysis

The key part of our method is the feature fusion based on triple attention, including the IOC type-level attention, the metapath-based neighbor node-level attention, and the metapath semantic-level attention. We conduct an explanatory analysis of our method from the perspective of triple attention fusion, which enables a better understanding of the importance of IOC types, metapath-based neighbor nodes, and metapaths through their corresponding attention weights.

**Analysis of IOC type-level attention.** Our method can learn attention weights that represent the importance of IOC types for report node feature completion and threat actor attribution. We take APT report R562[12] attributed to the Lazarus group as an example. R562 describes the investigation of an APT campaign that targeted the defense industry with the ThreatNeedle malware cluster. It includes numerous IOCs, which can be represented as 55 malware nodes, 44 URL nodes, 30 domain nodes, 30 technique nodes, seven filepath nodes, and two registry nodes. Figure 3 shows some representative neighbor nodes and relationships of R562. We sum the attention weights of the neighbor nodes belonging to the same IOC type and obtain the attention weight distribution for each IOC type, as shown in Figure 7a. The neighbor nodes of domains, URLs and malware play an important role in the report node feature completion of R562, while those of techniques, filepaths and registries contribute less. A further analysis indicates that many domain, URL and malware neighbors are specific to the Lazarus group, especially some malicious domains and URLs reported in multiple campaigns as critical indicators, significantly contributing to the attribution of R562. Our model captures these aspects and accordingly pays more attention to them. By contrast, many technique neighbors are more common across various APT groups; the filepath and registry neighbors are more amenable to change and less commonly reported; our model thereby pays less attention to them. In sum, the IOC type-level attention distinguishes between the nodes of different IOC types and focuses more on the important IOC types for threat actor attribution.

**Analysis of metapath-based neighbor node-level attention.** Our method can learn attention weights that represent the importance of metapath-based neighbors for heterogeneous feature fusion and threat actor attribution. We continue to take the example of APT report R562 and conduct the analysis based on metapath $MP_5$ (Report-Domain-Report), which indicates that two APT reports include the same domain. For $MP_5$, the metapath-based neighbors of R562 are three report nodes (i.e. R539[13], R528[14], and R284[15]), which describe multiple campaigns attributed to the Lazarus group. Their attention weights for R562 are shown in Figure 7b. Intuitively, the attention weights of these three report neighbors are not significantly different. R539 has more attention weights than the others, possibly because it has more common neighbors with R562.

---

[12] Kaspersky. Lazarus targets defense industry with ThreatNeedle. February 2021

[13] Google. New campaign targeting security researchers. January 2021

[14] Shusei Tomonaga. BLINDINGCAN - Malware Used by Lazarus -. September 2020

[15] HvS-Consulting. Greetings from Lazarus. December 2020

**Analysis of metapath semantic level attention.** Our method can learn attention weights that represent the importance of metapaths for threat actor attribution on the overall dataset. The metapaths with significant attention weights among the twenty designed metapaths are shown in Figure 7c. CTI reports usually contain much information about malware, which appears in different APT campaigns as different variants. Our model captures these aspects and provides much attention to metapath $MP_2$ (Report-Malware-Report) and metapath $MP_{13}$ (Report-Malware-Malware-Report). In addition, malicious domains connect different APT campaigns as important components of CTI reports and contribute greatly to threat actor attribution. Therefore, our model also pays significant attention to metapath $MP_5$ (Report-Domain-Report).

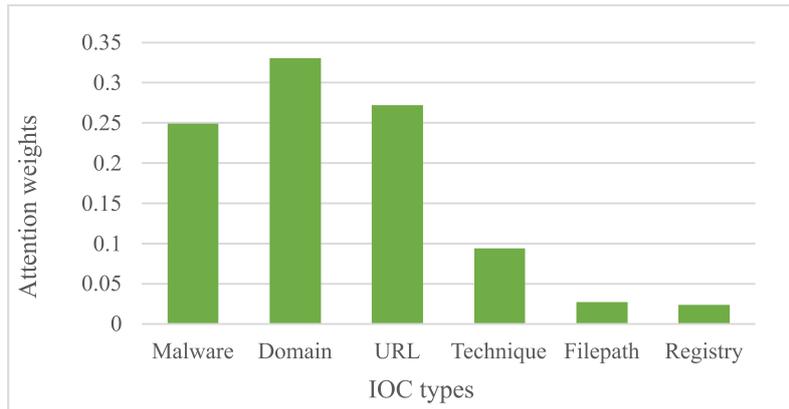

a. IOC type-level attention

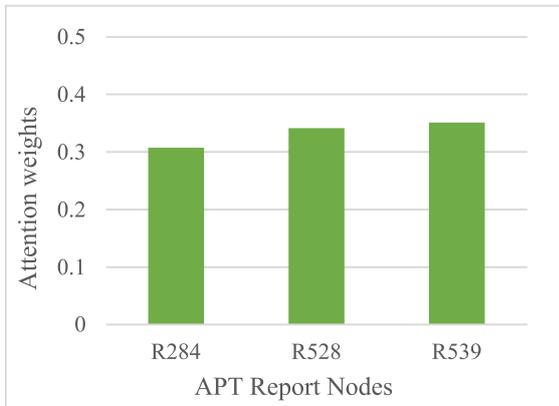

b. Metapath-based neighbor node-level attention based on $MP_5$

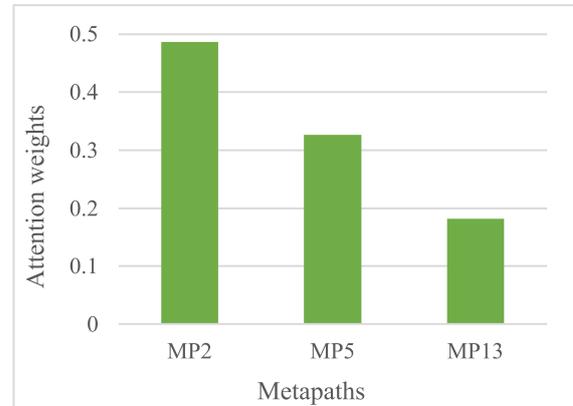

c. Metapath semantic-level attention

**Figure 7. R562-related triple attention weights**

## 5. Conclusion

In this paper, we address the insufficient feature extraction and fusion problems encountered in CTI-based APT actor attribution research and propose a method based on multimodal and multilevel feature fusion. The proposed APT-MMF method leverages a heterogeneous attributed graph to characterize APT reports and their IOC information. Our method extracts and fuses three types of modal features, i.e., attribute type features, natural language text features, and topological relationship features, to construct comprehensive node representations. APT-MMF further learns the deep hidden features of APT report nodes based on multilevel heterogeneous graph attention networks, including the IOC type-level, metapath-based neighbor node-level, and metapath semantic-level attention mechanisms. We construct a heterogeneous attributed graph dataset based on multisource threat intelligence for verification purposes. The experimental results show that our method is significantly better than the existing methods, achieving a Micro-$F_1$ value of 83.21% and a Macro-F1 value of 70.51% on multiclassification tasks concerning APT actor attribution. In addition, upon analyzing the triple attention mechanisms, the proposed APT-MMF approach demonstrates good interpretability for attribution analysis scenarios. In future work, we will mine more attribution-related features to improve the model and further study the identification methods of unknown APT actors.

# Acknowledgements

This work was supported by State Key Laboratory of Software Development Environment [grant number SKLSDE-2020ZX-02].

# References

[1] R. Ross. Managing Information Security Risk: Organization, Mission, and Information System View, Special Publication (NIST SP), National Institute of Standards and Technology, Gaithersburg, MD, 2011. https://tsapps.nist.gov/publication/get_pdf.cfm?pub_id=908030.

[2] Kaspersky Lab. Equation Group: Questions And Answers, 2015. https://media.kasperskydaily.com/wp-content/uploads/sites/92/2019/02/19083005/Equation_group_questions_and_answers.pdf.

[3] T. Rid, B. Buchanan. Attributing Cyber Attacks. Journal of Strategic Studies. 2015; 38(1–2):4–37. https://doi.org/10.1080/01402390.2014.977382.

[4] L. Chaoge, F. Binxing, L. Baoxu, C. Xiang, L. Qixu. A Hierarchical Model of Targeted Cyber Attacks Attribution. Journal of Cyber Security. 2019; 4(4):1–18.

[5] T. Steffens. Attribution of Advanced Persistent Threats: How to Identify the Actors Behind Cyber-Espionage, Springer, Berlin, Heidelberg, 2020. https://doi.org/10.1007/978-3-662-61313-9.

[6] Y. Mei, W. Han, S. Li, X. Wu, K. Lin, Y. Qi. A Review of Attribution Technical for APT Attacks. in: 2022 7th IEEE International Conference on Data Science in Cyberspace (DSC), 2022: p. 512–518. https://doi.org/10.1109/DSC55868.2022.00077.

[7] F. Skopik, T. Pahi. Under false flag: using technical artifacts for cyber attack attribution. Cybersecurity. 2020; 3(1):8. https://doi.org/10.1186/s42400-020-00048-4.

[8] S. Caltagirone, A. Pendergast, C. Betz. The diamond model of intrusion analysis. Threat Connect. 2013; 298(0704):1–61.

[9] C. Zhouguo, P. Shi, H. Yao, H. Chen. Levels Analysis of Network Attack Traceback. Computer Systems & Applications. 2014; 23(1):1–7.

[10] T. Pahi, F. Skopik. Cyber attribution 2.0: Capture the false flag. in: Proceedings of the 18th European Conference on Cyber Warfare and Security (ECCWS 2019), 2019: p. 338–345.

[11] A. Warikoo. The Triangle Model for Cyber Threat Attribution. Journal of Cyber Security Technology. 2021; 5(3–4):191–208. https://doi.org/10.1080/23742917.2021.1895532.

[12] V.S.C. Putrevu, H. Chunduri, M.A. Putrevu, S. Shukla. A Framework for Advanced Persistent Threat Attribution using Zachman Ontology. in: European Interdisciplinary Cybersecurity Conference, ACM,

Stavanger Norway, 2023: p. 34–41. https://doi.org/10.1145/3590777.3590783.

[13] P. Shakarian, G.I. Simari, G. Moores, S. Parsons. Cyber attribution: An argumentation-based approach. in: Cyber Warfare, Springer, 2015: p. 151–171.

[14] E. Nunes, P. Shakarian, G.I. Simari, A. Ruef. Argumentation models for cyber attribution. in: 2016 IEEE/ACM International Conference on Advances in Social Networks Analysis and Mining (ASONAM), 2016: p. 837–844. https://doi.org/10.1109/ASONAM.2016.7752335.

[15] E. Nunes, P. Shakarian, G. Simari. Toward argumentation-based cyber attribution. in: Workshops at the Thirtieth AAAI Conference on Artificial Intelligence, 2016.

[16] E. Karafili, L. Wang, A.C. Kakas, E. Lupu. Helping forensic analysts to attribute cyber-attacks: an argumentation-based reasoner. in: International Conference on Principles and Practice of Multi-Agent Systems, Springer, 2018: p. 510–518.

[17] E. Karafili, L. Wang, E.C. Lupu. An Argumentation-Based Reasoner to Assist Digital Investigation and Attribution of Cyber-Attacks. Forensic Science International: Digital Investigation. 2020; 32:300925. https://doi.org/10.1016/j.fsidi.2020.300925.

[18] S. Qamar, Z. Anwar, M.A. Rahman, E. Al-Shaer, B.-T. Chu. Data-driven analytics for cyber-threat intelligence and information sharing. Computers & Security. 2017; 67:35–58.

[19] Z. Zhu, R. Jiang, Y. Jia, J. Xu, A. Li. Cyber Security Knowledge Graph Based Cyber Attack Attribution Framework for Space-ground Integration Information Network. in: 2018 IEEE 18th International Conference on Communication Technology (ICCT), 2018: p. 870–874. https://doi.org/10.1109/ICCT.2018.8600108.

[20] Y. Ren, Y. Xiao, Y. Zhou, Z. Zhang, Z. Tian. CSKG4APT: A Cybersecurity Knowledge Graph for Advanced Persistent Threat Organization Attribution. IEEE Transactions on Knowledge and Data Engineering. 2022; 35(6):1–15. https://doi.org/10.1109/TKDE.2022.3175719.

[21] Y. Shen, G. Stringhini. ATTACK2VEC: Leveraging Temporal Word Embeddings to Understand the Evolution of Cyberattacks. in: 28th USENIX Security Symposium (USENIX Security 19), USENIX Association, Santa Clara, CA, 2019: p. 905–921. https://www.usenix.org/conference/usenixsecurity19/presentation/shen.

[22] A. Alsaheel, Y. Nan, S. Ma, L. Yu, G. Walkup, Z.B. Celik, X. Zhang, D. Xu. ATLAS: A sequence-based learning approach for attack investigation. in: 30th USENIX Security Symposium (USENIX Security 21), 2021: p. 3005–3022.

[23] T. Chen, C. Dong, M. Lv, Q. Song, H. Liu, T. Zhu, K. Xu, L. Chen, S. Ji, Y. Fan. APT-KGL: An Intelligent APT Detection System Based on Threat Knowledge and Heterogeneous Provenance Graph Learning. IEEE Transactions on Dependable and Secure Computing. 2022:1–15. https://doi.org/10.1109/TDSC.2022.3229472.

[24] X. Zang, J. Gong, X. Zhang, G. Li. Attack scenario reconstruction via fusing heterogeneous threat intelligence. Computers & Security. 2023; 133:103420. https://doi.org/10.1016/j.cose.2023.103420.

[25] F.K. Kaiser, U. Dardik, A. Elitzur, P. Zilberman, N. Daniel, M. Wiens, F. Schultmann, Y. Elovici, R. Puzis. Attack Hypotheses Generation Based on Threat Intelligence Knowledge Graph. IEEE Transactions on Dependable and Secure Computing. 2023; 20(6):4793–4809. https://doi.org/10.1109/TDSC.2022.3233703.

[26] V. Sachidananda, R. Patil, A. Sachdeva, K.Y. Lam, L. Yang. APTer: Towards the Investigation of APT Attribution. in: 2023 IEEE Conference on Dependable and Secure Computing (DSC), 2023: p. 1–10. https://doi.org/10.1109/DSC61021.2023.10354155.

[27] Y. Qiao, X. Yun, Y. Zhang. How to Automatically Identify the Homology of Different Malware. in:

2016 IEEE Trustcom/BigDataSE/ISPA, 2016: p. 929–936. https://doi.org/10.1109/TrustCom.2016.0158.

[28] S. Li, Q. Zhang, X. Wu, W. Han, Z. Tian. Attribution Classification Method of APT Malware in IoT Using Machine Learning Techniques. Security and Communication Networks. 2021; 2021:9396141. https://doi.org/10.1155/2021/9396141.

[29] Y. Mei, W. Han, S. Li, K. Lin. A hybrid intelligent approach to attribute Advanced Persistent Threat Organization using PSO-MSVM Algorithm. IEEE Transactions on Network and Service Management. 2022; 19(4):4262–4272. https://doi.org/10.1109/TNSM.2022.3201928.

[30] I. Rosenberg, G. Sicard, E. David. DeepAPT: Nation-State APT Attribution Using End-to-End Deep Neural Networks. in: Artificial Neural Networks and Machine Learning, Springer International Publishing, 2017: p. 91–99.

[31] Q. Wang, H. Yan, Z. Han. Explainable APT Attribution for Malware Using NLP Techniques. in: 2021 IEEE 21st International Conference on Software Quality, Reliability and Security (QRS), 2021: p. 70–80. https://doi.org/10.1109/QRS54544.2021.00018.

[32] G. Laurenza, L. Aniello, R. Lazzeretti, R. Baldoni. Malware Triage Based on Static Features and Public APT Reports. in: Cyber Security Cryptography and Machine Learning, Springer International Publishing, 2017: p. 288–305. https://doi.org/10.1007/978-3-319-60080-2_21.

[33] J. Liu, Y. Shen, H. Yan. Functions-based CFG Embedding for Malware Homology Analysis. in: 2019 26th International Conference on Telecommunications (ICT), 2019: p. 220–226. https://doi.org/10.1109/ICT.2019.8798769.

[34] O. Mirzaei, R. Vasilenko, E. Kirda, L. Lu, A. Kharraz. SCRUTINIZER: Detecting Code Reuse in Malware via Decompilation and Machine Learning. in: Detection of Intrusions and Malware, and Vulnerability Assessment, Springer International Publishing, 2021: p. 130–150. https://doi.org/10.1007/978-3-030-80825-9_7.

[35] U. Noor, Z. Anwar, T. Amjad, K.K.R. Choo. A machine learning-based FinTech cyber threat attribution framework using high-level indicators of compromise. Future Generation Computer Systems. 2019; 96:227–242. https://doi.org/10.1016/j.future.2019.02.013.

[36] Y. Shin, K. Kim, J.J. Lee, K. Lee. ART: Automated Reclassification for Threat Actors based on ATT&CK Matrix Similarity. in: 2021 World Automation Congress (WAC), 2021: p. 15–20. https://doi.org/10.23919/WAC50355.2021.9559514.

[37] I. Lee, C. Choi. Camp2Vec: Embedding cyber campaign with ATT&CK framework for attack group analysis. ICT Express. 2023; 9(6):1065–1070. https://doi.org/10.1016/j.icte.2023.05.008.

[38] E. Irshad, A. Basit Siddiqui. Cyber threat attribution using unstructured reports in cyber threat intelligence. Egyptian Informatics Journal. 2022; 26(1):43–59. https://doi.org/10.1016/j.eij.2022.11.001.

[39] L. Perry, B. Shapira, R. Puzis. NO-DOUBT: Attack Attribution Based On Threat Intelligence Reports. in: 2019 IEEE International Conference on Intelligence and Security Informatics (ISI), 2019: p. 80–85. https://doi.org/10.1109/ISI.2019.8823152.

[40] N. S, R. Puzis, K. Angappan. Deep Learning for Threat Actor Attribution from Threat Reports. in: 2020 4th International Conference on Computer, Communication and Signal Processing (ICCCSP), 2020: p. 1–6. https://doi.org/10.1109/ICCCSP49186.2020.9315219.

[41] K.Z. Huang, Y.F. Lian, D.G. Feng, H.X. Zhang, D. Wu, X.L. Ma. Method of Cyber Attack Attribution Based on Graph Model. Journal of Software. 2022; 33(2):683–698.

[42] B. Tang, J. Wang, Z. Yu, B. Chen, W. Ge, J. Yu, T. Lu. Advanced Persistent Threat intelligent

profiling technique: A survey. Computers and Electrical Engineering. 2022; 103:108261. https://doi.org/10.1016/j.compeleceng.2022.108261.

[43] Y. Guo, Z. Liu, C. Huang, N. Wang, H. Min, W. Guo, J. Liu. A framework for threat intelligence extraction and fusion. Computers & Security. 2023; 132:103371. https://doi.org/10.1016/j.cose.2023.103371.

[44] N. Sun, M. Ding, J. Jiang, W. Xu, X. Mo, Y. Tai, J. Zhang. Cyber Threat Intelligence Mining for Proactive Cybersecurity Defense: A Survey and New Perspectives. IEEE Communications Surveys & Tutorials. 2023; 25(3):1748–1774. https://doi.org/10.1109/COMST.2023.3273282.

[45] X. Zhao, R. Jiang, Y. Han, A. Li, Z. Peng. A survey on cybersecurity knowledge graph construction. Computers & Security. 2024; 136:103524. https://doi.org/10.1016/j.cose.2023.103524.

[46] J. Devlin, M.-W. Chang, K. Lee, K. Toutanova. BERT: Pre-training of Deep Bidirectional Transformers for Language Understanding. in: Proceedings of the 2019 Conference of the North, Association for Computational Linguistics, Minneapolis, Minnesota, 2019: p. 4171–4186. https://doi.org/10.18653/v1/N19-1423.

[47] A. Grover, J. Leskovec. node2vec: Scalable Feature Learning for Networks. in: Proceedings of the 22nd ACM SIGKDD International Conference on Knowledge Discovery and Data Mining, 2016: p. 855–864.

[48] X. Wang, H. Ji, C. Shi, B. Wang, Y. Ye, P. Cui, P.S. Yu. Heterogeneous Graph Attention Network. in: The World Wide Web Conference, ACM, New York, NY, USA, 2019: p. 2022–2032. https://doi.org/10.1145/3308558.3313562.

[49] V. Mavroeidis, S. Bromander. Cyber Threat Intelligence Model: An Evaluation of Taxonomies, Sharing Standards, and Ontologies within Cyber Threat Intelligence. in: 2017 European Intelligence and Security Informatics Conference (EISIC), IEEE, 2017: p. 91–98. https://doi.org/10.1109/EISIC.2017.20.

[50] OASIS. STIX 2.1. 2021; https://oasis-open.github.io/cti-documentation/.

[51] B.E. Strom, A. Applebaum, D.P. Miller, K.C. Nickels, A.G. Pennington, C.B. Thomas. MITRE ATT&CK: Design and Philosophy, 2020. https://attack.mitre.org/resources/.

[52] MITRE. Common Vulnerabilities and Exposures. 2023; https://cve.mitre.org.

[53] S. Sebastián, J. Caballero. AVclass2: Massive Malware Tag Extraction from AV Labels. in: Annual Computer Security Applications Conference, Association for Computing Machinery, New York, NY, USA, 2020: p. 42–53. https://doi.org/10.1145/3427228.3427261.

[54] VirusTotal. VirusTotal. 2023; . https://virustotal.com.

[55] B. Perozzi, R. Al-Rfou, S. Skiena. Deepwalk: Online learning of social representations. in: Proceedings of the 20th ACM SIGKDD International Conference on Knowledge Discovery and Data Mining, 2014: p. 701–710.

[56] Y. Dong, N.V. Chawla, A. Swami. Metapath2vec: Scalable representation learning for heterogeneous networks. in: Proceedings of the ACM SIGKDD International Conference on Knowledge Discovery and Data Mining, 2017: p. 135–144. https://doi.org/10.1145/3097983.3098036.

[57] T.N. Kipf, M. Welling. Semi-Supervised Classification with Graph Convolutional Networks. in: 5th International Conference on Learning Representations, International Conference on Learning Representations, ICLR, 2016. http://arxiv.org/abs/1609.02907.

[58] P. Veličković, G. Cucurull, A. Casanova, A. Romero, P. Liò, Y. Bengio. Graph Attention Networks. in: 6th International Conference on Learning Representations, 2017. http://arxiv.org/abs/1710.10903.

[59] D. Jin, C. Huo, C. Liang, L. Yang. Heterogeneous Graph Neural Network via Attribute Completion.

in: Proceedings of the Web Conference 2021, ACM, Ljubljana Slovenia, 2021: p. 391–400. https://doi.org/10.1145/3442381.3449914.

[60] Y. Gao, L.I. Xiaoyong, P. Hao, B. Fang, P. Yu. Hincti: A cyber threat intelligence modeling and identification system based on heterogeneous information network. IEEE Transactions on Knowledge and Data Engineering. 2020; 34(2):708–722.